\documentclass[useAMS,usenatbib]{mn2e}

\usepackage{fixltx2e,ifpdf,amsmath,amssymb}
\ifpdf
  \usepackage[pdftex]{graphicx,color}
\else
  \usepackage[dvips]{graphicx,color}
\fi

\newcommand{\Teff}{\ensuremath{T_{\rm eff}}}
\newcommand{\logg}{\ensuremath{\log{g}}}

\newcommand{\vt}{\ensuremath{v_{\rm t}}}
\newcommand{\kms} {\mbox{\rm km$\;$s$^{-1}$}}
\newcommand{\atlas}{\mbox{\sc Atlas}}
\newcommand{\atlasIX}{\mbox{{\sc Atlas}9}}
\newcommand{\atlasXII}{\mbox{{\sc Atlas}12}}
\newcommand{\ang}{\mbox{\AA}}
\newcommand{\logten}{\ensuremath{\log_{10}}}

\newcommand{\strom}{Str\"{o}mgren}
\newcommand{\pr}{\ensuremath{\phantom{\mbox{r}}}}
\title[Limb-darkening coefficients]
{New limb-darkening coefficients and synthetic photometry for
  model-atmosphere grids at Galactic, LMC, and SMC abundances.}
\author
[Ian D. Howarth]
{Ian D. Howarth\thanks{E-mail: idh@star.ucl.ac.uk}\\
Dept.\ Physics  \& Astronomy, UCL, Gower Street, London
WC1E~6BT, UK}
\begin{document}

\date{Accepted. Received; in original form}

\pagerange{\pageref{firstpage}--\pageref{lastpage}} \pubyear{2010}

\maketitle

\label{firstpage}

\begin{abstract}
  New grids of \atlasIX\ models have been calculated using revised
  convection parameters and updated opacity-distribution functions,
  for chemical compositions intended to be representative of solar,
  [M/H] = +0.3, +0.5, Large Magellanic Cloud (LMC), and Small
  Magellanic Cloud (SMC) abundances.  The grids cover $\Teff =
  3.5{\mbox{--}}50$kK, from $\logg = 5.0$ to the effective Eddington
  limit.  Limb-darkening coefficients and synthetic photometry are
  presented in the \mbox{\emph{UBVRIJHKLM}}, \emph{uvby},
  \emph{ugriz}, WFCAM, Hipparcos/Tycho, and Kepler passbands for these
  models, and for Castelli's comparable `new-ODF' grids.  Flux
  distributions are given for the new models.  The sensitivity of
  limb-darkening coefficients to the adopted physics is illustrated.
\end{abstract}

%\begin{keywords}
%circumstellar matter -- infrared: stars.
%\end{keywords}

\section{Introduction}

The significance of limb darkening as a probe of atmospheric structure
was recognized from the earliest modelling of stellar atmospheres
\citep[e.g.,][]{Schwarz06, Jeans17, Milne21}, and its importance in
the quantitative photo\-metric analysis of eclipsing binaries similarly
noted \citep{Russell12, RusShp12}.  Reliable empirical determinations
of limb darkening are rarely possible in eclipsing binaries (in part because
of degeneracies with other parameters), and so values derived from
model-atmosphere calculations retain their importance through to the
present day, where new applications include modelling of
exoplanetary transits, microlensing
events, and spatially resolved stellar surfaces
\citep[e.g.,][]{Southworth08, Witt95, Hestroffer97, Aufdenberg05}.

Wholesale calculation of limb-darkening coefficients became feasible
only after production of the first large grids of model atmospheres
\citep[cf.][who review earlier studies]{Grygar72}.  In particular, the
extensive model-atmosphere grids generated by \citet{Kurucz79,
  Kurucz93} form the basis of work by \citet{Wade85}, \citet{Diaz92},
\citet{vanHamme93}, \citet{Claret00}, and others.  

Although several programs are now available in the public domain for
generating line-blanketed model atmospheres which relax the
approximation of Local Thermodynamic Equilibrium (LTE), Kurucz's
\atlas\ codes remain the benchmark for LTE calculations, and have two
major attractions for the present purposes.  First, while the models
may not accurately reproduce individual spectral lines for which
non-LTE effects are significant, they nonetheless appear to be quite
successful in generating accurate atmospheric structures and
broad-band fluxes.

Secondly, it is straightforward, and computationally cheap, to
generate self-consistent \atlas\ grids covering large ranges in
parameters of interest.  The aspect of internal consistency is of
particular importance when modelling the spectra of systems whose
temperatures and gravities may show considerable variations over their
surfaces (such as the components of close binary systems, rapidly
rotating stars, and non-radial pulsators), and when attempting uniform
analyses of samples encompassing a range of stellar parameters (for
example, in synoptic studies of light-curves of exoplanetary transits).

Almost all extensive calculations of grids of \atlas-based
limb-darkening coefficients published to date have used the
formulations of opacity-distribution functions (ODFs) and atmospheric
convection inherent to the models published by
\citet{Kurucz93}.\footnote{An exception is 
work reported by \citet{Barban03}}  However, 
improved treatements in both areas have emerged subsequently, leading
in some cases to quite significant changes in the models
\citep{Castelli04}.
The principal purpose of the present paper is to present 
calculations of limb-darkening coefficients made using new ODFs and
improved parameterizations of convection.   Because of growing
interest in modelling eclipsing binaries in the Magellanic Clouds 
\citep[e.g.,][]{Harries03, Hilditch05, Bonanos09, North10}, new models
with tailored MC abundances are included.

\section{Models}

\atlas\ has undergone more or less continuous development over its
$\sim$40-year lifetime, and currently exists in two distinct forms,
\atlasIX\ (which uses opacity-distribution functions) and \atlasXII\
(which uses opacity sampling).
The models discussed here were all computed using the Trieste port of 
\atlasIX\ to {\sc gnu}-linux systems \citep{Sbordone07}.

\subsection{ODFs and Abundances}

Opacity-distribution functions are required as a basic input to
\atlasIX\ models.  \citet{Castelli04} describe ODFs which incorporate a
variety of improvements over those used in the \citet{Kurucz93} grids,
of which perhaps the most important is the inclusion of improved and
additional molecular opacities.  The models described here use new
ODFs, generated using
{\sc dfsynthe} 
\citep{Kurucz05, Castelli05},
 based on the \citet{Castelli04} line lists.
%%calculated for five sets of abundances (and the
%%four microturbulent velocities which are standard in this context).

The baseline models use solar abundances reported by
\citet{Asplund05}.  Limb darkening plays an important role in the
analysis of light-curves of stars with transiting exoplanets, and the
sample of such stars currently known shows a propensity for enhanced
metallicity \citep[e.g.,][]{Gonzalez97, Fischer05, Sousa08}.  With
this in mind, additional models (and ODFs) were calculated with
metallicities enhanced by +0.3 and +0.5 dex over solar values.  For
test purposes, ODFs (but not model grids) were computed at several
other metallicities, including [M/H] = $-0.5$ for the Vega models
discussed in Section~\ref{sec:phot1}.

The routine study of eclipsing binaries in external galaxies, and
especially in the Magellanic Clouds (MCs), has become feasible in
recent years.  While no one set of MC abundances is likely to be
generally satisfactory, most interest in eclipsing binaries in the
Clouds has concentrated on early-type stars (in part because of their
intrinsic brightness).  In order to have an appropriate (Pop.~I)
reference set of representative Large and Small Magellanic Cloud
abundances that are at least well defined, and apparently reasonable,
results have been compiled from a number of sources.
These abundances,
listed in Appendix~\ref{sec:Appx1} (Table~\ref{tab:MCab}),
have been used to compute new LMC and SMC ODFs.

All the ODFs newly calculated here are available on-line.

\subsection{Convection}

All model-atmosphere codes incorporate approximations and
parameterizations of physics that may be difficult or expensive to model, or is
simply poorly understood \citep[cf., e.g.,][]{Kurucz96}.  Convection is
a prime example, and is characterized in \atlasIX\ by a mixing-length
approximation with optional overshooting.

The widely distributed \citet{Kurucz93} grids were calculated with a
ratio of mixing length to pressure scale height of $\ell/H = 1.25$ and overshooting
included, but \citet{Castelli97} argue that more-consistent results
are obtained with \atlasIX\ when overshooting is switched off.
Furthermore, \citet[][and personal communication]{Smalley05} advocates
smaller values of the mixing-length parameter over at least the late-A
spectral range.  

All the models presented here were calculated with no overshooting. The baseline grids
use $\ell/H = 1.25$, but supplementary grids at $\ell/H = 0.50$ were also computed.

\begin{figure}
\center{\includegraphics[scale=0.38,angle=-90]{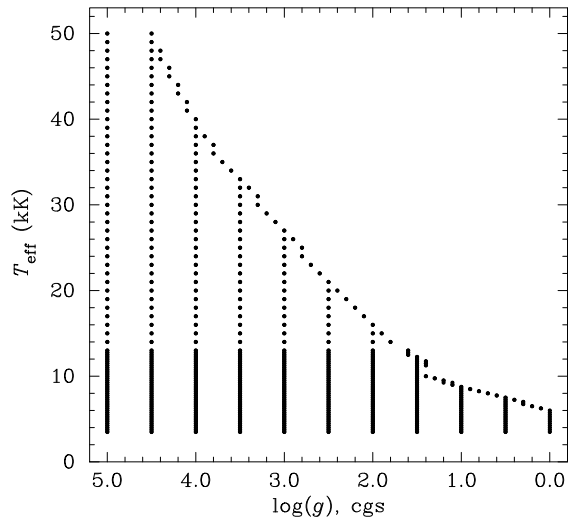}} 
\caption{Grid for solar-abundance models
(the lowest gravities may be
0.1--0.2~dex larger at higher metallicities).}
\label{fig:grid}
\end{figure}

\begin{table*}
  \caption{Summary of \atlasIX\ models.  
The sources for
the atmospheric structures are \citeauthor{Castelli04}
(\citeyear{Castelli04}; \atlasIX.C04) and the present paper
(\atlasIX.A10).
In the `Grid' columns,
with entries of the form `xIIvJJ', `xII' indicates the scaling applied to metals
from the default solar abundance (such that `m10' indicates 
a metallicity [M/H]$ = -1.0$, and `p05' +0.5), while `vJJ' indicates the
turbulent velocity used in the model (`JJ' in \kms).  Models at LMC
and SMC abundances are so labelled.    The quantity
$\ell/H$ is
ratio of mixing length to pressure scale height.
The number of models in a given grid, $N$, is greater for the \atlasIX.A10
set because these generally extend to somewhat lower gravities 
than the \atlasIX.C04 calculations, at all 76 grid temperatures.
}
  \begin{tabular}{ccccccccc}
  \hline
Structure source&&Grid &$\ell/H$ &$N$ &$\;$ &Grid & $\ell/H$ & $N$ \\
 \hline
\atlasIX.C04 && m25v02  &1.25 &476  &&m40av02  &1.25 &476 \\
\atlasIX.C04 && m20v02  &1.25 &476  &&m25av02  &1.25 &476 \\
\atlasIX.C04 && m15v02  &1.25 &476  &&m20av02  &1.25 &476 \\
\atlasIX.C04 && m10v02  &1.25 &476  &&m15av02  &1.25 &476 \\
\atlasIX.C04 && m05v02  &1.25 &476  &&m10av02  &1.25 &476 \\
\atlasIX.C04 && p00v00  &1.25 &476  &&m05av02  &1.25 &476 \\
\atlasIX.C04 && p00v02  &1.25 &476  &&p00av02  &1.25 &476 \\
\atlasIX.C04 && p02v02  &1.25 &476  &&p02av02  &1.25 &476 \\
\atlasIX.C04 && p05v02  &1.25 &476  &&p05av02  &1.25 &476 \\
\\
\atlasIX.A10 &&P00v00  &1.25 &554   &&LMCv02  &1.25 &554\\
\atlasIX.A10 &&P00v02  &1.25 &554   &&{\pr}LMCv02r &0.50 &552\\
\atlasIX.A10 &&{\pr}P00v02r &0.50 &552   &&SMCv02  &1.25 &554\\
\atlasIX.A10 &&P00v04  &1.25 &548   &&{\pr}SMCv02r &0.50 &552\\
\atlasIX.A10 &&P03v02  &1.25 &543\\
\atlasIX.A10 &&{\pr}P03v02r &0.50 &544\\
\atlasIX.A10 &&P05v02  &1.25 &531\\
\atlasIX.A10 &&{\pr}P05v02r &0.50 &533\\
\hline
\label{tab:Grids}
\end{tabular}
\end{table*}

\subsection{Parameter space}

The grid sampling adopted here matches 
the basic $\Teff/\logg$ sampling of the \citet{Castelli04} models; that is,
from 3.5--13kK at 0.25kK intervals, and 12.0--50.0kK at 1kK intervals,
from $\logg = 5.0$~dex~(cgs) to lower values at 0.5-dex intervals.
This grid density is sufficient for interpolation of output products
(e.g., linearly in $\log{\Teff}/\logg$) to be satisfactory for most
applications.  Some effort has been put into generating models to as
low a gravity as possible at a given temperature (the effective
Eddington limit), and so the present
grids extend beyond \citeauthor{Castelli04}'s in this regard.
The sampling is illustrated in
Fig.~\ref{fig:grid}.

Previous work suggests that the \atlasIX\ turbulent-velocity parameter $\vt$
has little consequence for limb-darkening in the optical regime,
and its value (as well as its physical significance) is difficult to
establish observationally in any individual case.  Most of the model
intensities presented here
have been computed with a canonical $\vt = 2$~\kms.
Solar-abundance grids at $\vt = 0$ and 4~\kms\ confirm the
insensitivity of limb darkening to {\vt}.

Results of grids of 6571 new models are presented here.
\citet{Castelli04} have calculated models using solar
abundances listed by \citet{Grevesse98}, which differ only slightly
from those given by \citet{Asplund05}.  Their grids extend over a
greater range in scaled solar abundances than those newly calculated
here, and include additional modifications to abundances of
alpha-process elements.  For completeness, intensities have also been
calculated from the 8568 atmospheric structures they provide, along
with limb-darkening coefficients and other products to match the new
grids.  Table~\ref{tab:Grids} summarizes the grids, where `\atlasIX.C04'
and `\atlasIX.A10' identify the \citet{Castelli04} and current grids, respectively.

\section{Data Products}

The principal motivation for the work reported here was to provide
up-to-date limb-darkening coefficients (particularly at MC abundances). 
Thus the main data products are specific intensities
(monochromatic radiances) $I_\lambda(\mu)$, 
as a function of $\mu$, the cosine of
the angle between the line of sight and the surface normal.
These are
provided in condensed form as broad-band limb-darkening coefficients.

Physical fluxes,
\begin{equation}
\begin{split}
F_\lambda &= 2\pi \int_0^1{I_\lambda(\mu) \mu \,\mbox{d}\mu}\\
& = 4 \pi H_\lambda
\end{split}
\label{eq:flux}
\end{equation}
(where $H_\lambda$ is the Eddington flux), are required as a check on
the accuracy of intensity integrations, and in any case represent only
a minor additional computational overhead.  Model fluxes are therefore
also provided, both at the 1221 wavelengths used in the structure
calculations (91{\ang}--160$\mu$m), and in broad-band (`synthetic
photometry') form.  Other data products (such as model structures and
higher-resolution intensities) will be made freely available on
request.

\section{Limb Darkening}

\subsection{Passbands}
\label{sec:bands}

Modern detectors working in the optical are essentially photon-counting in nature.
For such detectors
broad-band model intensities (and, {\em{mutatis mutandis,}} fluxes)
may be calculated as
\begin{equation}
I_i(\mu) =
\frac
{\int_\lambda{ I_\lambda(\mu)  \phi_{\lambda}(i)  \lambda \mbox{d}\lambda }}
{\int_\lambda{\phi_{\lambda}(i) \lambda \mbox{d}\lambda }
},
\end{equation}
(e.g., \citeauthor{Bessell05} \citeyear{Bessell05})
where $\phi_{\lambda}(i)$ is the response function of 
passband $i$ in a given
photo\-metric system.  
Historically, the equivalent formalism for energy-integrating detectors
has been used:
\begin{equation}
I_i(\mu) =
\frac
{\int_\lambda{ I_\lambda(\mu)  \phi_{\lambda}(i)  \mbox{d}\lambda }}
{\int_\lambda{\phi_{\lambda}(i) \mbox{d}\lambda }.
}
\end{equation}

Results are provided for photon-counting detectors for all photometric
systems considered here, and additionally
for energy-integrating detectors for the Johnson and \strom\ passbands.
The adopted sources for response functions are:
\begin{itemize}
\item 
\emph{UBVRI} (Johnson--Cousins system): \citet{Bessell90}.
\item
\emph{JHKLL$^\prime$M} (Johnson--Glass system): \citet{Bessell88}.
\item
\emph{uvby} (Str\"{o}mgren system):  \citeauthor{Crawford70} (\citeyear{Crawford70};
filters),
\citeauthor{Stubbs07} (\citeyear{Stubbs07};  atmospheric
transmission),
\citeauthor{Hamamatsu99} (\citeyear{Hamamatsu99};  S4 response),
\citeauthor{Allen76} (\citeyear{Allen76}, p.~108;  aluminium reflectivity),
\citeauthor{pgo} (BK7 glass transmission).\footnote{The \strom\ system is close to being filter defined, but atmospheric
transmission, 1P21-S4 photomultiplier sensitivity, and BK7 glass
transmission all make roughly equal inroads into the short-wavelength edge of the
$u$ passband.}
\item
\emph{ZYJHK} (WFCAM system): \citet{Hewett06}
\item
\emph{H$_p$, B$_T$, V$_T$} (Hipparcos/Tycho system): \citet{Bessell00}
\item
\emph{ugriz} (Sloan system): on-line,\newline 
\texttt{http://www.sdss3.org/instruments/camera.php}
\item
\emph{Kepler}: on-line,\newline 
{\tt{http://keplergo.arc.nasa.gov/kepler\_response\_hires1.txt}}
\end{itemize}

Simple geometric treatements of mutual irradiation (the `reflection
effect')
in binary-star systems 
require
the angular dependence of the bolometric intensity
\citep{Wilson90}, so bolometric
limb-darkening coefficients have also been calculated ($\phi_\lambda
\equiv 1$).

\subsection{Analytical characterization}
\label{sec:CLD}

Specific intensities
are traditionally, and conveniently,
characterized by simple functional forms, of which the most
venerable is the linear limb-darkening law
\begin{equation}
I(\mu)/I(1)  = 1 - u(1-\mu),
\label{eq01}
\end{equation}
the analytical solution for
a source function that is linear in optical depth $\tau$
(\citealt{Schwarz06}, \citealt{Milne21};
$u=0.6$ for a grey atmosphere).

More-realistic models do not have analytical solutions, and additional
terms are required for a faithful functional representation of actual
limb darkening.  \citet{Kopal49} took a series-expansion approach, and
found the differences between quadratic and quartic approximations to
be negligible at the level of accuracy with which he was concerned.
Two-coefficient parameterizations subsequently became standard; in
particular, a quadratic law of the form
\begin{equation}
I(\mu)/I(1) = 1 - a(1-\mu) - b(1-\mu)^2
\label{eq02}
\end{equation}
was widely adopted in the modern computational era
\citep[e.g.,][]{Manduca77,Wade85},
although a cubic version,
\[
I(\mu)/I(1) = 1 - a(1-\mu) - b(1-\mu)^3, \nonumber
\]
was advocated by \citet{vantVeer60}.
An alternative logarithmic law, 
\begin{equation}
I(\mu)/I(1) = 1 - a(1-\mu) - b(\mu\ln{\mu})
\label{eq05}
\end{equation}
was proposed by \citet{Klinglesmith70}, 
and a square-root law,
\begin{equation}
I(\mu)/I(1) = 1 - a(1-\mu) - b(1-\sqrt{\mu}),
\label{eq04}
\end{equation}
by \citet{Diaz92}.

For modern work, these limb-darkening laws suffer from various
limitations.
The obvious way to develop a single analytical expression that
more accurately reproduces numerical limb-darkening results across a
wide parameter space is simply to use more terms, a notion
implemented by \citet{Claret00} who introduced
a four-coefficient fit in powers of $\sqrt\mu$,
\begin{equation}
I(\mu)/I(1) =  1 -
\sum_{n=1}^4a_n\left({1 - \mu^{n/2}}\right).
\label{eq06}
\end{equation}

\begin{figure}
\center{\includegraphics[scale=0.30,angle=-90]{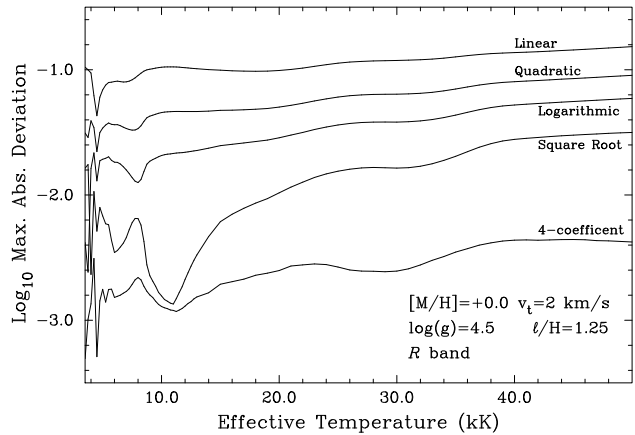}} 
\caption{Comparison of different parameterizations of limb darkening
(equations~\ref{eq01}--\ref{eq06}).
The quantity
plotted is $\log_{10}$ of the maximum absolute difference between
input and parameterized values of $I(\mu)/I(1)$.}
\label{Fig1}
\end{figure}

\begin{figure}
\center{\includegraphics[scale=0.30,angle=-90]{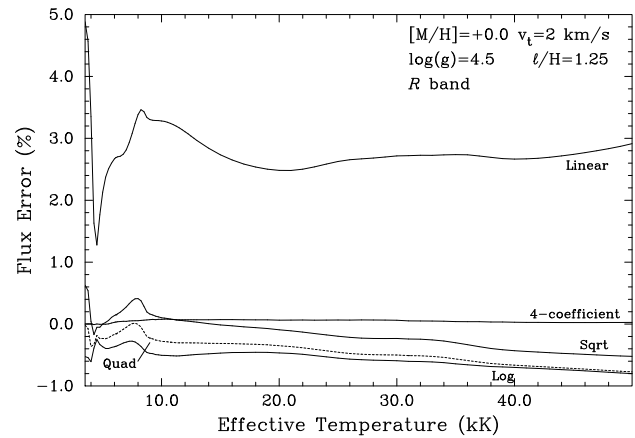}} 
\caption{Differences, flux minus integrated intensity (expressed as a
  percentage of flux) for
  several limb-darkening laws (see Section~\ref{sec:FP}).}
\label{Fig2}
\end{figure}

Figure~\ref{Fig1} illustrates the differing degrees to which the
foregoing common functional forms reproduce the intensities computed
here.  While quantitative results depend on the chosen normalization
and adopted figure of merit, the qualitative conclusions are robust: a
linear limb-darkening law is a rather poor representation of detailed
numerical results; a quadratic law fares little better; but the
four-coefficient law affords an improvement of up to an order of
magnitude in precision over any standard two-coefficient option.  While there
has been some discussion in the literature as to which is the most
appropriate limb-darkening `law' to adopt for light-curve analyses,
and in particular under circumstances which permit empirical
determination of limb-darkening coefficients
\citep[e.g.,][]{Southworth08}, there seems little justification for
adopting anything other than a high-order fit such as eqtn.~(\ref{eq06})
if the principal purpose is to secure an accurate representation of
model-atmosphere results.

\subsection{Fitting procedure}
\label{sec:FP}

For any given analytical form, there remains a choice of numerical
technique for determining the limb-darkening coefficients.  The most
direct approach is simple least-squares determination of coefficients
from functional fits to computed $I(\mu)$ values (with $\hat{I}(1)$,
the fitted value of $I(1)$, as an optional additional free parameter).
A related approach, the $r$-integration method \citep{Heyrovsky07},
appears to offer no significant benefits \citep{Claret08}.

A criticism that can be levelled against the least-squares method is
that subsequent integration of parameterized intensities may not
accurately recover the flux, as it should (eqtn.~\ref{eq:flux}).
Alternative formulations in which which flux is explicitly conserved
can be devised \citep[e.g.,][]{Wade85, vanHamme93}, giving rise to
so-called flux conservation methods \citep[cf., e.g.,][for a recent
discussion]{Claret08}.  Such methods require some further {\em ad hoc}
constraint; e.g., \citet{Wade85} set $\hat{I}(1) \equiv I(1)$ for the
linear model.

For the most part, straightforward least-squares fitting has been used
here, to intensities evaluated at 20 angles.\footnote{For comparison
  purposes, the linear law has been modelled with and without $I(1)$
  as an optimized parameter, and with explicit flux conservation;
  Appendix~\ref{sec:Appx2}.}  Subsequent analytical integration of the
four-coefficient intensity model then reproduces directly computed
fluxes to better than 0.1\%.  When fit in this manner, all
two-coefficient models conserve flux to within $\sim$1\%, while the
linear model again fares poorly (Fig.~\ref{Fig2}).

\begin{table*}
  \caption{Johnson colour-index normalizations.   Column 2 gives the colour zero-point computed from
the Vega 9.55kK model, with photometry adopted from \citeauthor{Bessell98} 
(\citeyear{Bessell98};  e.g., observed $(V-R) = -0.009$).  Column 3 gives the equivalent results for Sirius.
The recommended zero-points, $Z_{P12}$, are the average of the
previous two columns (with corrections for alternative photometries
dropped);
synthetic colours may then be computed from model fluxes by using eqtn.~(\ref{eq:CI}).
\newline
$C_E$ is the correction to be added to columns~2--4 for energy-integrating detectors, and $C_{9.4}$ is the correction to be added to column~2
to obtain results for the 9.40kK Vega model.
The $U_X$ and $B_X$ passbands are those recommended by 
\citet{Bessell90} for use in computing synthetic colours to be compared with
observed $(U-B)$ colours.}
  \begin{tabular}{cllccccccccccccccccccc}
  \hline   
   Colour  &\multicolumn{1}{c}{Vega 9.40kK}
                                        &\multicolumn{1}{c}{Sirius 9.85kK}      
                                                                    &$ Z_{P12}$&$   C_E  $&$   C_{9.4} $\\
\hline
$(U_X-B_X)$&$  -0.422 +[(U-B) - 0.000] $&$ -0.418 +[(U-B) + 0.045] $&$ -0.420 $&$ +0.031 $&$ +0.020$\\
$  (B-V)  $&$  +0.609 +[(B-V) - 0.000] $&$ +0.608 +[(B-V) + 0.010] $&$ +0.608 $&$ +0.000 $&$ +0.007$\\
\\
$  (V-R)  $&$  +0.568 +[(V-R) + 0.009] $&$ +0.575 +[(V-R) + 0.010] $&$ +0.572 $&$ +0.018 $&$ +0.005$\\
$  (V-I)  $&$  +1.260 +[(V-I)\,\, + 0.005] $&$ +1.276 +[(V-I)\,\, + 0.016] $&$ +1.268 $&$ -0.002 $&$ +0.010$\\
$  (V-K)  $&$  +4.924 +[(V-K) - 0.020] $&$ +4.918 +[(V-K) + 0.061] $&$ +4.921 $&$ -0.002 $&$ +0.026$\\
\\
$  (J-K)  $&$  +2.261 +[(J-K) - 0.010] $&$ +2.249 +[(J-K) + 0.018] $&$ +2.255 $&$ -0.002 $&$ +0.006$\\
$  (H-K)  $&$  +1.156 +[(H-K)]         $&$ +1.152 +[(H-K) + 0.009] $&$ +1.152 $&$ +0.002 $&$ +0.002$\\
$  (K-L)  $&$  +1.872 +[(K-L) - 0.010] $&$ +1.872 +[(K-L) - 0.003] $&$ +1.872 $&$ -0.001 $&$ +0.002$\\
\hline
\label{tab:zp}
\end{tabular}
\end{table*}

\section{Photometric matters}
\label{sec:phot1}

Broad-band fluxes have potential utility in a
variety of applications (effective-temperature determinations,
population synthesis, etc.).  Thus while it is not the intention here
to conduct a detailed confrontation of the new model fluxes with
observations, a basic transformation of the synthetic photometry to
observed magnitude systems may be of interest.  Modern systems
generally have dedicated programmes of flux calibration, so the focus
here is on the Johnson and \strom\ systems.

\subsection{Johnson system}
\subsubsection{Flux zero-point}

Full normalization of the synthetic photometry can be divided into two
parts:  absolute flux calibration, and differential (colour) calibration.
The former aspect is traditionally bound to
the absolute flux calibration of Vega, which has been
addressed in detail by \citet{Hayes85} and
\citet{Megessier95}, using primary measurements mostly made in the
1970s.
Their key results can be summarized as 
pseudo-monochromatic 5556-\ang\ fluxes:
\begin{align}
f(5556) =   &\   3.44 \times 10^{-9} 
&&\mbox{erg cm}^{-2}\mbox{ s}^{-1}\mbox{\ang}^{-1}\nonumber\\
&\ 3.46\times 10^{-11} 
&&\mbox{W m}^{-2}\mbox{nm}^{-1}\nonumber
\end{align}
respectively.
These results can be used to provide a zero-point calibration for the
present grids by using a model matching Vega, and its observed
$V$-band magnitude.

Both `the' $V$-band magnitude, and the choice of model, are subject to
uncertainty; widely adopted values are $V=0.03$ \citep{Johnson66}, and
$\Teff = 9550$K, $\logg=3.95$, $\mbox{[M/H]} = -0.5$, $\vt = 2$~\kms\
\citep{Castelli94}.  Subsequent analyses have supported this effective
temperature \citep[e.g.][]{Ciardi01, Hill10}, but the HST calibration
programme uses a model with $\Teff = 9400$K, $\logg=3.90$,
$\mbox{[M/H]} = -0.5$, $\vt = 0$~\kms\ \citep{Bohlin07}.  Workers are
therefore faced with choices in observed flux, observed magnitude, and
adopted model, as well as in whether to adopt energy-integrating or
photon-counting magnitudes.  The situation is further complicated by
the fact that Vega is a rapid rotator viewed pole-on \citep{Gray88,
Gulliver91, Aufdenberg06, Hill10}, so that in principle no
single-temperature model can give an accurate representation of the
entire flux distribution (although for magnitude zero-point
calibration this is of minor importance, because the models are only
used to scale between 5556-\ang\ and $V$-band fluxes).

The baseline adopted here is the Hayes calibration, 9.55kK model,
\citet{Johnson66} $V$, and photon-counting magnitudes.   For
this system the
$V$-band flux 
$f(V)$ is 
$1.031 \times f(5556)$,
yielding
\begin{equation}
\begin{split}
V = &-2.5\log_{10}f(V) - 21.096\\
    &+C_1 + C_2 + C_3+ C_4
\label{eq:Vband}
\end{split}
\end{equation}
where 
$f(V)$ is in 
erg cm$^{-2}$ s$^{-1}$  \ang$^{-1}$ and
the optional correction factors are\newline
$C_1 = -0.0004$ for the $\Teff = 9400$~K model;\newline
$C_2 = +0.0063$ for the \citet{Megessier95} $f(5556)$ flux; \newline
{\pr\pr}more generally, $C_2 = 2.5\log_{10}[f(5556)/f(\mbox{Hayes})]$;\newline
$C_3= +0.0135$ for energy-integrated photometry;  and\newline
$C_4= V_{\rm Vega} - 0.03$.

\subsubsection{Colour-index zero-points}

Differential normalization relies on
relating the observed colours of a standard star, or stars, to 
model fluxes.  A model colour index is formed from fluxes in
two passbands $P1$ and $P2$
by
\begin{align}
(P1 - P2)_{\rm M} & = -2.5\log_{10}\left[{
\frac{f(P1)}{f(P2)} }\right] + (C_{P1}- C_{P2}) \nonumber\\
& \equiv -2.5\log_{10}\left[{
\frac{f(P1)}{f(P2)} }\right] + Z_{P12}
\label{eq:CI}
\end{align}
where $f(P1), f(P2)$ are model broad-band fluxes and
$C_{P1}, C_{P2}, Z_{P12}$ are constants.
Equating the right-hand side of eqtn.~(\ref{eq:CI})
with the observed (reddening-free) colour index
yields the required normalizing constant $Z_{P12}$.  

Vega has traditionally been adopted as the primary calibrating star, with
increasinging use of Sirius as a supplementary standard since the work
of \citet{Cohen92}.   Once again, however, observed photometry is
subject to both statistical uncertainties and
deterministic changes (e.g., one might choose to {\em define} Vega to have
colour indexes of zero, or to average values from an ensemble of
standard stars), and the appropriate model parameters are also moot.
Here the photometry compiled by \citet{Bessell98} is adopted, as are
the Sirius 
model parameters due to Kurucz which they quote:
$\Teff = 9.85$kK, $\logg = 4.25$, [M/H] = +0.5, $\vt = 2$~\kms\
(see also \citeauthor{Castelli99} \citeyear{Castelli99}).   These
parameters are in good agreement with other recent determinations (e.g., 
\citealt{Hill93}; \citealt{vanNoort98}; \citealt{Qiu01}).

The resulting normalizing constants are listed in Table~\ref{tab:zp}.
Agreement between the Vega and Sirius calibrations for the normalizing
constants inferred from the specific set of observed colours adopted here
is better for the 9.55kK Vega model, and the adopted
$Z_{P12}$ values are the average of this model and the Sirius results.

\subsubsection{Bolometric corrections}

From the basic definitions of bolometric correction and magnitude, it is straightforward to show that the bolometric correction in some passband $P$
for a model at effective temperature \Teff\ is given by
\begin{align}
BC(P) &= M_{{\rm Bol, }\odot} 
-2.5\log_{10}\left[{
\frac{d_{10}^2}{T_\odot^4R_\odot^2} }\right]\nonumber \\
&-2.5\log_{10}\Teff^4       
+2.5\log_{10}(f_P) +C_P
\label{eq:bolo}
\end{align}
where 
$d_{10}$ is 10~pc (expressed in the same units as the solar radius, $R_\odot$) and 
$C_P$ is the magnitude-normalizing constant, which can be found for the Johnson passbands
from data summarized in the preceding two 
subsections.\footnote{E.g.,
$C_B \equiv (C_B - C_V) + C_V = +0.608 -21.096 = -20.488$,\newline
where numerical values for photon-counting photometry are  from
 Table~\ref{tab:zp} and
eqtn.~(\ref{eq:Vband}).}    

Bolometric corrections in the main Johnson bands are included for each
model in the on-line listings, adopting $M_{{\rm Bol, }\odot} =
+4.74$.  As a check on results, a solar-like model was calculated
with $\Teff = 5778$K, $\logg = 4.44$~(cgs), $\vt = 2$~\kms, $\ell/H =
1.25$, $\mbox{[M/H]} = 0.0$; for this model, the (photon-counting)
$V$-band bolometric correction is found to be $BC(V) = -0.075$, in
excellent agreement with the empirical solar value of $-0.07$
\citep{Bessell98}.

\begin{table*}
  \caption{Summary of \strom\ photometric calibrations (see Section~\ref{sec:calS}). The second column gives flux-calibration and colour zero-points
for photon-counting detectors, using a $\Teff = 9.55$kK model,
\citet{Crawford70} photometry and the \citet{Hayes85} flux calibration for Vega.\newline
\mbox{$\quad$}$C_1$ is the correction to be added to the values in column~2 if adopting other sources of photometry;\newline
\mbox{$\quad$}$C_2$ is the correction to be added for energy-integrating detectors;\newline
\mbox{$\quad$}$C_3$ is the correction to be added for the \citet{Megessier95} flux calibration.\newline
Column 6 gives the corrections to be added for a $\Teff = 9.40$kK Vega model, while the final column 
gives the corrections for Sirius photometry (baseline data from
\citealt{CrawfordBG70}) and a 9.85kK model, where  $\Delta{V} \equiv
V_{\rm Vega} - 0.03$.}
  \begin{tabular}{cclcccclcc}
  \hline
Band/    &Vega       &   \multicolumn{1}{c}{$C_1$}     
                                        &   $C_2$ &  $C_3$  &&  Vega     &\multicolumn{1}{c}{Sirius}\\
Colour  &   9.55kK  &               &          &          & &  9.40kK   &\multicolumn{1}{c}{9.85kK}\\
  \hline
$u$     & $-19.784$ & $+(u-1.445)$  & $+0.001$ &$+0.0063$ & & $-0.045$ &$+\Delta{V} - 0.037 + (u-1.298)$\\
$v$     & $-20.147$ & $+(v-0.195)$  & $+0.000$ &$+0.0063$ & & $-0.016$ &$+\Delta{V} - 0.019 + (v-0.166)$\\
$b$     & $-20.573$ & $+(b-0.034)$  & $+0.001$ &$+0.0063$ & & $-0.011$ &$+\Delta{V} + 0.012 + (b-0.036)$\\
$y$     & $-21.074$ & $+(V-0.03) $  & $+0.002$ &$+0.0063$ & & $-0.006$ &$+\Delta{V} - 0.002$\\
\\
$(b-y)$ & $-0.500$  &$+0.004 - (b-y)$&$+0.000$ &          & & $+0.004$ &$-0.015 + [0.006-(b-y)]$\\
$m_1$ & $+0.074$  & $+0.157 - m_1$&$+0.001$ &          & & $+0.001$ &$+0.047 + [0.124-m_1]$\\
$c_1$ & $+0.064$  & $+1.089 - c_1$&$-0.002$ &          & & $+0.023$ &$-0.015 + [1.002 - c_1]$\\
\hline
\end{tabular}
\label{tab:stromZ}
\end{table*}

\subsection{A Str\"omgren calibration}
\label{sec:calS}

Flux calibration is readily carried out for the \strom\ \emph{uvby}
system in the same way as for the Johnson system.  A potential
complication is that for the intermediate-width passbands of this
system, individual spectral features may be significant; in
particular, H$\delta$ (410nm) falls in the middle of the $v$ band.  To
check that the standard 20-\ang\ spectral sampling is adequate in
these circumstances, high-resolution synthetic spectra were computed
from the Vega models.  \strom-band fluxes from these synthetic spectra
agree with the standard results to better than 0.1\%\ for \emph{u, b,}
and \emph{y}, and are only 0.5\%\ smaller at \emph{v}.  Since the
Balmer lines reach their greatest strength at early-A spectral types,
the standard sampling appears to be satisfactory.

Results are summarized in Table~\ref{tab:stromZ}.   
\strom\ colours can be generated from the models with
\begin{align}
(b-y) &= -2.5\logten\left[{  
\frac{f(b)}{f(y)}  
}\right] 
+ Z_{by}\nonumber\\
m_1 &= -2.5\logten\left[{ \frac{f(v)\times f(y)}{f^2(b)}}\right] + Z_{m1}\nonumber\\
c_1 &= -2.5\logten\left[{ \frac{f(u)\times f(b)}{f^2(v)}}\right] + Z_{c1}\nonumber
\end{align}
where $f(p)$ is the model flux in passband $p$ and the zero-points $Z$
are the last three entries in column~2 of Table~\ref{tab:stromZ}.

\citet{Fabregat96} and \citet{Gray98} have previously provided \strom\
flux calibrations which are the results of convolving the \emph{uvby}
passbands with observed spectrophotometry,\footnote{The Gray
  \emph{v}-band calibration is model-based.} from \citet{Glushneva92}
and \citet{Taylor84}, respectively (both nominally on the
\citealt{Hayes85} system).  Comparison with their results, expressed
as a zero-magnitude flux in units of $10^{-9}$ erg cm$^{-2}$ s$ ^{-1}$
\ang$^{-1}$ for consistency with their presentations, shows generally
good agreement:
\begin{center}
\begin{tabular}{cccclcclcc}
Band   & FR  &Gray & Atlas9.A10 \\
$u$    &11.80 & 11.72 & 12.20  \\
$v$    &8.69  & (8.66) & 8.73\\
$b$    &5.84  &  5.89  & 5.90\\
$y$    & 3.69 & 3.73 & 3.72\\
\end{tabular}
\end{center}

\noindent
The $\sim$4\%\ difference at \emph{u} between spectrophotometric and
model results may be a consequence of uncertainties in observed Vega
magnitudes, which range over 10\%\ in this passband \citep{Perry69,
  Barry69, JG70, Crawford70}; e.g., using Sirius as the calibrator yields a
zero-magnitude \emph{u}-band flux of $11.79 \times 10^{-9}$ erg
cm$^{-2}$ s$ ^{-1}$ \ang$^{-1}$.

\section[]{Discussion}

Much attention has been paid in the literature as to which is the best
analytical representation of limb darkening, and which is the
appropriate numerical technique for evaluating limb-darkening
coefficients (see discussion and references in
Sections~\ref{sec:CLD},~\ref{sec:FP}).  While these are important
details, the use of a high-order parameterization, which provides
satisfactory results under most circumstances, renders such details
academic.

\begin{figure}
\center{\includegraphics[scale=0.30,angle=-90]{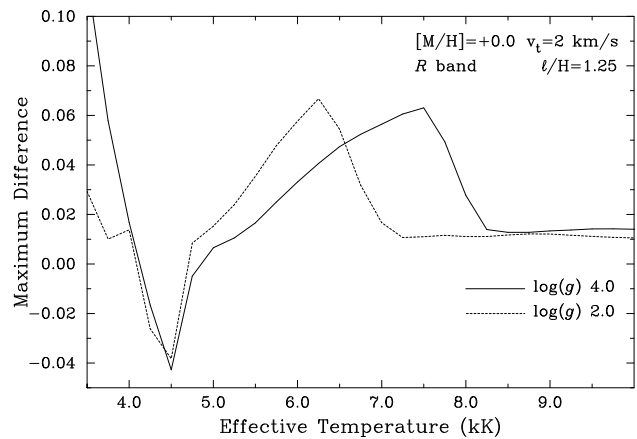}} 
\caption{Maximimum differences in $I(\mu)/I(1)$
between
the results of the present models and fits by \citet{Claret00} to
earlier Kurucz models.}
\label{Fig3}
\end{figure}

\begin{figure}
\center{\includegraphics[scale=0.55,angle=-90]{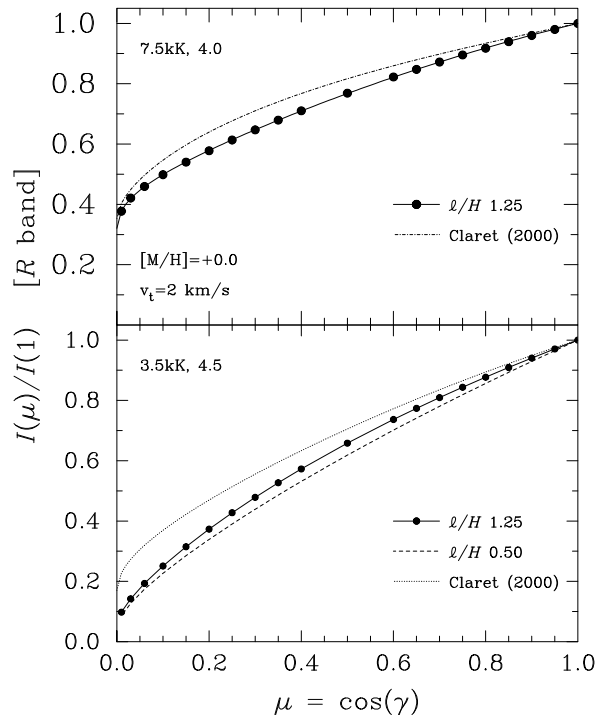}} 
\caption{Normalized $R$-band intensities 
from the present calculations for two $\Teff/\log{g}$ values.
Filled circles are calculated intensities, and  
continuous lines the corresponding four-coefficient fits.
The \citet{Claret00} results characterize earlier \atlasIX\ models.}
\label{Fig4}
\end{figure}

However, the larger importance of the
underlying model atmospheres has not generally been emphasized,
perhaps because most analysts have used `off-the-shelf'
model-atmosphere results.  
The new opacity sources introduced in the \citet{Castelli04} models
show up directly in emergent fluxes and intensities, and the
changes in treatment of convection also change the structures.
These changes are particularly important at $\Teff \lesssim 8$kK;
for models at $\Teff \gtrsim 10$kK the
differences between the current models, those of \citet{Castelli04},
and the older \citet{Kurucz93} grids, are small, and
result largely from revisions to adopted abundances.   

As far as specific intensities are concerned, differences between the
newer calculations and older results are illustrated by a comparison
between the extensive listings provided by \citet{Claret00} and the
\atlasIX.A10 models presented here (Fig.~\ref{Fig3}).  The changes can be much
larger than any that arise through different fitting techniques,
provided that the adopted functional form is adequate
(Fig.~\ref{Fig4}).

We might expect further developments in \atlas\ models in future, but
there is reason to hope that consequences may be modest in scale, at
least in the short term; using the latest version of his
opacity-sampling \atlasXII\ code, Kurucz (personal communication) has
kindly computed a solar-abundance model at $\Teff = 7.5$kK, $\logg =
4.0$, $\vt = 2$~\kms. Differences between old and new \atlasIX\ models
are relatively large for these parameters, but further changes in the
\atlasXII\ products are negligible for the present purposes.

Finally, the limitations of the present work should be recognised.
First, although the Kurucz/Castelli ODFs used here represent a
significant improvement over older compilations, they are still
inevitably incomplete.  In particular, there are a number of
indications that \atlasIX\ models begin to show significant
disagreement with observations for effective temperatures
$\lesssim$4kK (e.g., \citealt{Bessell98, Bertone08}), presumably
because of the increasing importance of missing molecular opacities.
Models at the coolest effective temperatures should be used with caution.

Secondly, the synthetic photometry can be expected to show not only
the successes in reproducing observations
reported for similar models by \citet{Bessell98} and
\citet{Castelli99}, but also the failures;  note that
\citet[][cf.\ their Appendix
E.3.1]{Bessell98} find that their modelled $(U-B)$ colours (i.e., using
$U_X- B_X$ passbands) required an ad hoc scaling of $0.96\times$ to
bring them into agreement with observations.

Thirdly, it must be owned that the new models presented here
fare no better than old ones in matching the handful of apparently
reliable empirical determinations of stellar limb darkening
\citep[e.g.,][]{Heyrovsky07,Claret08,Claret09}, although such
determinations are usually limited to estimates of coefficients for
linear or, occasionally, quadratic limb-darkening laws for stars other
than the Sun (and such laws are poor representations of model
results).  Common simplifying model-atmosphere approximations for scattering phase
functions will be a factor in this; although such approximations may
be appropriate for structure and integrated-flux calculations, they
have obvious weaknesses for evaluating intensities (or polarizations).
Addressing these issues is  a matter for future study.

\section{Recommendations}

Practically all modern studies in the optical region
use photon-counting detectors, and so models appropriate to that
context will normally be used.
The investigator is still faced with choices in model convection,
microturbulence, and abundance (Table~\ref{tab:Grids}), and in
limb-darkening parameterization (equations~\ref{eq01}--\ref{eq06}).
If, as is frequently the case, there are no results from detailed
analyses to suggest otherwise, it's reasonable to adopt the
appropriate global abundance system (solar, LMC, or SMC), $\vt =
2$~\kms, $\ell/H = 1.25$.  The careful investigator might well conduct
sensitivity tests to examine the consequences of other parameters
(including varying $\Teff$ and $\log{g}$); in particular, for
effective temperatures less than $\sim$8kK, it would be prudent to
examine the effects of models computed at $\ell/H = 0.50$.

Other than for comparison with empirically-determined limb-darkening
laws (normally linear,  but increasingly for quadratic coefficients),
there is little merit in adopting anything other than a
four-coefficient limb-darkening parameterization (equation~\ref{eq06})
if seeking an accurate representation of model-atmosphere intensities.
The computational cost compared to simpler formulae is trivial in
anything other than Monte-Carlo calculations, and even there the
overhead should be entirely tolerable.

\section*{Acknowledgments}

I gratefully acknowledge the contribution made by R.L.~Kurucz in
making his codes open source and his datasets freely available, and
the work carried out by F.~Castelli and her colleagues in porting them
to run on {\sc gnu}-linux systems.  I thank Susana Barros, Mike
Bessell, Antonio Claret, Steve Fossey, Urtzi Jauregi, Bob Kurucz, Luca Sbordone,
Barry Smalley, and the referee for helpful correspondence and comments.

\onecolumn
\appendix

\section[]{Adopted Magellanic-Cloud Abundances}
\label{sec:Appx1}

Baseline
solar abundances were taken from \citet{Asplund05}. Cloud 
abundances were adopted, in order
of preference, from: FLAMES results (\citealt{Hunter07}, \citealt{Trundle07}, as
summarized by \citealt{Evans08}); \citet{Garnett99}; and \citet[][applying their
differential values to Asplund solar abundances]{Russell92}.   In each
case, samples of early-type stars or H$\;${\sc ii} regions were analysed, so the
abundances are representative of Population~I.
For elements not studied in these sources,
solar abundances were used, adjusted by the median metal offsets from \citet{Russell92}:
$-0.3$ (LMC) and $-0.6$~dex (SMC).  These global adjustments are in
adequate agreement with more-recent discussions 
\citep[e.g.,][]{Hill95, Hill97, Luck98, Venn99, Mokiem07}.
\begin{table*}
  \caption{Adopted LMC and SMC abundances, by number, on a scale where
    the hydrogen abundance is 12.0 dex.}
  \begin{tabular}{rlrrrrlrrrrlrrrrlrr}
 \hline
\multicolumn{2}{c}{Element}&  \multicolumn{1}{c}{LMC}  &  \multicolumn{1}{c}{SMC} &$\;$&
\multicolumn{2}{c}{Element}&  \multicolumn{1}{c}{LMC}  &  \multicolumn{1}{c}{SMC} &$\;$&
\multicolumn{2}{c}{Element}&  \multicolumn{1}{c}{LMC}  &  \multicolumn{1}{c}{SMC} &$\;$&
\multicolumn{2}{c}{Element}&  \multicolumn{1}{c}{LMC}  &  \multicolumn{1}{c}{SMC}  \\
 \hline
 2 &He & 10.88 & 10.85&& 24 & Cr&  5.43 &  5.06&& 46 & Pd&  1.39 &  1.09 && 68 & Er&  0.63 &  0.33 \\
 3 &Li &  0.75 &  0.45&& 25 & Mn&  5.06 &  4.88&& 47 & Ag&  0.64 &  0.34 && 69 & Tm&$-0.30$&$-0.60$\\
 4 &Be &  1.08 &  0.78&& 26 & Fe&  7.23 &  6.93&& 48 & Cd&  1.47 &  1.17 && 70 & Yb&  0.78 &  0.48 \\
 5 &B  &  2.40 &  2.10&& 27 & Co&  4.62 &  4.32&& 49 & In&  1.30 &  1.00 && 71 & Lu&$-0.24$&$-0.54$\\
 6 &C  &  7.73 &  7.37&& 28 & Ni&  6.02 &  5.82&& 50 & Sn&  1.70 &  1.40 && 72 & Hf&  0.58 &  0.28 \\
 7 &N  &  6.90 &  6.50&& 29 & Cu&  3.91 &  3.63&& 51 & Sb&  0.70 &  0.40 && 73 & Ta&$-0.47$&$-0.77$\\
 8 &O  &  8.35 &  7.98&& 30 & Zn&  4.20 &  4.00&& 52 & Te&  1.89 &  1.59 && 74 & W &  0.81 &  0.51 \\
 9 &F  &  4.26 &  3.96&& 31 & Ga&  2.58 &  2.28&& 53 & I &  1.21 &  0.91 && 75 & Re&$-0.07$&$-0.37$\\
10 &Ne &  7.60 &  7.20&& 32 & Ge&  3.28 &  2.98&& 54 & Xe&  1.97 &  1.67 &&76  & Os&  1.15 &  0.85 \\
11 &Na &  6.97 &  5.78&& 33 & As&  1.99 &  1.69&& 55 & Cs&  0.77 &  0.47 &&77  & Ir&  1.08 &  0.78 \\
12 &Mg &  7.06 &  6.72&& 34 & Se&  3.03 &  2.73&& 56 & Ba&  1.94 &  1.23 &&78  & Pt&  1.34 &  1.04 \\
13 &Al &  6.07 &  6.27&& 35 & Br&  2.26 &  1.96&& 57 & La&  1.06 &  0.84 &&79  & Au&  0.71 &  0.41 \\
14 &Si &  7.19 &  6.79&& 36 & Kr&  2.98 &  2.68&& 58 & Ce&  1.45 &  1.26 &&80  & Hg&  0.83 &  0.53 \\
15 &P  &  5.06 &  4.76&& 37 & Rb&  2.30 &  2.00&& 59 & Pr&  0.41 &  0.11 &&81  & Tl&  0.60 &  0.30 \\
16 &S  &  6.70 &  6.30&& 38 & Sr&  2.46 &  1.32&& 60 & Nd&  1.66 &  1.47 &&82  & Pb&  1.70 &  1.40 \\
17 &Cl &  5.01 &  4.95&& 39 & Y &  1.88 &  1.60&&    &   &       &       &&83  & Bi&  0.35 &  0.05 \\
18 &Ar &  6.20 &  5.90&& 40 & Zr&  2.20 &  1.93&& 62 & Sm&  0.98 &  1.13 &&    &   &       &       \\
19 &K  &  4.78 &  4.48&& 41 & Nb&  1.12 &  0.82&& 63 & Eu&  0.22 &  0.25 &&90  & Th&$-0.24$&$-0.54$\\
20 &Ca &  5.83 &  5.63&& 42 & Mo&  1.62 &  1.32&& 64 & Gd&  0.82 &  0.52 &&    &   &       &       \\
21 &Sc &  2.62 &  2.31&&    &   &       &      && 65 & Tb&$-0.02$&$-0.32$&&92  & U &$-0.82$&$-1.12$\\
22 &Ti &  4.63 &  4.31&& 44 & Ru&  1.54 &  1.24&& 66 & Dy&  0.84 &  0.54 \\
23 &V  &  4.10 &  3.56&& 45 & Rh&  0.82 &  0.52&& 67 & Ho&  0.21 &$-0.09$\\
\hline
\end{tabular}
\label{tab:MCab}
\end{table*}

\newpage
\section{On-line material}
\label{sec:Appx2}

New opacity-distribution functions using the \citet{Asplund05} solar
abundances are provided on-line, at metallicities [M/H] = $-0.5, +0.0,
+0.2, +0.3, +0.4, +0.5$, together with the LMC and SMC ODFs.  In each
case, ODFs are given sampled at the standard `big' and `little'
wavelength sampling intervals required for use with \atlasIX, and for
microturbulent velocities of 0, 1, 2, 4, and 8~\kms, together with the
corresponding Rosseland-opacity files.   These are binary files in the
format required by \atlasIX.

Limb-darkening and flux data products are organized into directories
corresponding to the grids listed in Table~\ref{tab:Grids} (for both
new models, and those of \citealt{Castelli04}).  Within a given
directory, there are three plain-text files for each model, with names
of the form {\tt{t05000g40.<ext>}} where the base identifies \Teff\
and \logg, and the extension identifies the file content.

Files with {\tt{.flx}} extensions contain listings of physical fluxes
at 1221 wavelengths, from 90\ang\ to 160$\mu$m (with 20-\ang\ sampling
through the optical).  Fluxes are tabulated in erg cm$^{-2}$ s$^{-1}$
\ang$^{-1}$ as a function of wavelength (in \ang).

Limb-darkening coefficients are provided in files with extensions
{\tt{.ucP}} (for photon-counting detectors;  Section~\ref{sec:bands})
and {\tt{.ucE}} (for energy-integrating detectors).
The {\tt{.ucE}} files list  broad-band limb-darkening
coefficients for \strom\ and Johnson systems  (including $U_X$ and $B_X$ passbands for
computing synthetic $(U-B)$ colours), while the {\tt{.ucP}} files
give broad-band limb-darkening
coefficients for all 30 passbands listed in
Section~\ref{sec:bands}, plus [energy-integrated] bolometric
coefficients.
The data given for each passband may be 
illustrated by $V$-band results for a Vega model ($\Teff = 9.55$kK,
$\logg = 3.95$, [M/H]$ = -0.5$, $\vt = 2$~\kms, $\ell/H = 1.25$):
\begin{verbatim}
Bessell-V             5467.7  5.61840E+07  2.08372E+07    -0.207
  4-coeff       +4.73679E-01 +5.53030E-01 -5.01375E-01 +1.56613E-01     +2.75266E-04 -7.09332E-04
  quadratic     +2.25617E-01 +3.71954E-01                               +1.56919E-02 -4.19850E-02
  quad - FC     +2.74255E-01 +3.01889E-01                               +1.82965E-02 -6.25047E-02
  square root   +2.25306E-02 +6.82355E-01                               +2.77154E-03 +6.52661E-03
  logarithmic   +6.36370E-01 +3.25055E-01                               +5.45636E-03 -1.48612E-02
  linear - 2    +1.04831E+00 +5.92481E-01                               +3.77170E-02 -9.16531E-02
  linear - 1    +5.25808E-01                                            +4.55362E-02 -1.09348E-01
  linear - FC   +4.25199E-01                                            +7.68496E-02 -2.08951E-01
\end{verbatim}
Here the first line lists 
\begin{itemize}
\item[(i)] the passband, 
\item[(ii)] effective wavelength (in \ang),
\item[(iii)] the broad-band physical flux $f(\lambda)$ 
(in erg cm$^{-2}$ s$^{-1}$ \ang$^{-1}$), 
\item[(iv)] the broad-band surface-normal intensity $I_\lambda(1)$
(in erg cm$^{-2}$ s$^{-1}$ \ang$^{-1}$ sr$^{-1}$), and
\item[(v)] the bolometric correction (for Johnson passbands only).
\end{itemize}
Subsequent lines list coefficients for the limb-darkening laws
summarized in Section~\ref{sec:CLD};   the final two numbers 
in each row are the
r.m.s.\ and
maximum  differences between input and modelled values of
$I(\mu)/I(1)$.

The 
`{\tt{linear - FC}}' and
`{\tt{quad - FC}}'
results use flux conservation in place of least squares,
with the fitted value $\hat{I}(1)$ anchored to the model
$I(1)$ in the first case, and additionally
$\hat{I}(0.1)$ to $I(0.1)$ in the second (cf.\ \citealt{Wade85}).   The 
`{\tt{linear - 1}}' listing is for the standard (single-parameter)
least-squares
fit to eqtn.~(\ref{eq01}), while the coefficients in the 
`{\tt{linear - 2}}' listing are for a model with
$\hat{I}(1)$ free, i.e.,
\begin{equation}
I(\mu)/I(1)  = a - b(1-\mu).\nonumber
\end{equation}

\bsp

\label{lastpage}

\end{document}